\documentclass[%
twocolumn,
reprint,
groupedaddress,
 amsmath,amssymb,
 aip,
 jcp,
]{revtex4-1}

\usepackage{graphicx,tabularx}
\usepackage{xspace}
\usepackage{dcolumn}
\usepackage{bm}
\usepackage[version=3]{mhchem}
\usepackage{verbatim}
\usepackage[colorlinks=true, urlcolor=blue, citecolor=blue, filecolor=blue, linkcolor=blue]{hyperref}

\def\ket#1{\mathinner{|{#1}\rangle}} 
\newcommand{\wn}{\ensuremath{\mathrm{cm}^{-1}}\xspace}

\begin{document}

\title{Perspective: Tipping the scales - search for drifting constants from molecular spectra}

\author{Paul Jansen}
\author{Hendrick L. Bethlem}
\author{Wim Ubachs}
\affiliation{Department of Physics and Astronomy, LaserLaB, VU University Amsterdam,
De Boelelaan 1081, 1081 HV Amsterdam, The Netherlands}

\date{\today}

\begin{abstract}
Transitions in atoms and molecules provide an ideal test ground for constraining or detecting a possible variation of the fundamental constants of nature. In this Perspective, we review molecular species that are of specific interest in the search for a drifting proton-to-electron mass ratio $\mu$. In particular, we outline the procedures that are used to calculate the sensitivity coefficients for transitions in these molecules and discuss current searches. These methods have led to a rate of change in $\mu$ bounded to $6 \times 10^{-14}$/yr from a laboratory experiment performed in the present epoch. On a cosmological time scale the variation is limited to $|\Delta\mu/\mu| < 10^{-5}$ for look-back times of 10-12 billion years and to $|\Delta\mu/\mu| < 10^{-7}$ for look-back times of 7 billion years. The last result, obtained from high-redshift observation of methanol, translates into $\dot{\mu}/\mu = (1.4 \pm 1.4) \times 10^{-17}$/yr if a linear rate of change is assumed.
\end{abstract}

\maketitle

\section{Introduction\label{sec:introduction}}
The fine-structure constant, $\alpha \approx 1/137$, which determines the overall strength of the electromagnetic force, and the proton-to-electron mass ratio, $\mu=m_p/m_e \approx 1836$, which relates the strengths of the forces in the strong sector to those in the electro-weak sector\cite{Flambaum2004}, are the only two dimensionless parameters that are required for the description of the gross structure of atomic and molecular systems~\cite{Born1935}. The values of these two constants ensure that protons are stable, that a large number of heavy elements could form in the late evolution stage of stars, and that complex molecules based on carbon chemistry exist~\cite{Hogan2000}. If these constants would have had only slightly different values, even by fractions of a percent, our Universe would have looked entirely different. The question whether this \emph{fine tuning} is coincidental or if the constants can be derived from a -- yet unknown -- theory beyond the Standard Model of physics, is regarded as one of the deepest mysteries in science. One solution to this enigma may be that the values of the fundamental constants of nature may vary in time, or may obtain different values in distinct parts of the (multi)-Universe. Searches for drifting constants are  motivated by this perspective.


Theories that predict spatial-temporal variations of $\alpha$ and $\mu$ can be divided into three classes. The first class comprises a special type of quantum field theories that permit variation of the coupling strengths. Bekenstein postulated a scalar field for the permittivity of free space; this quintessential field then compensates the energy balance in varying $\alpha$ scenarios to accommodate energy conservation as a minimum requirement for a theory~\cite{Bekenstein1982}. Based on this concept various forms of dilaton theories with coupling to the electromagnetic part of the Lagrangian were devised, combined with cosmological models for the evolution of matter (including dark matter) and dark energy under the assumptions of General Relativity. Such scenarios provide a natural explanation for variation of fundamental constants over cosmic history, i.e., as a function of red-shift parameter $z$. The variation will freeze out under conditions, where the dark energy content has taken over from the matter content in the Universe, a situation that has been reached almost completely~\cite{Sandvik2002}. These theories provide a rationale for searches of drifting constants at large look-back times toward the origin of the Universe, even if laboratory experiments in the modern epoch were to rule out such variations. The second class of theories connects drifting constants to the existence of high order dimensions as postulated in many versions of modern string theory~\cite{Aharony2000}. Kaluza-Klein theories, first devised in the 1920s, showed that formulations of electromagnetism in higher dimensions resulted in different effective values of $\alpha$, after compactification to the four observed dimensions. Finally, the third class of theories, known as Chameleon scenarios, postulate that additional scalar fields acquire mass depending on the local matter density \cite{Khoury2004}.

Experimental searches for temporal variation of fundamental constants were put firmly on the agenda of contemporary physics by the ground-breaking study by Webb \emph{et al.}\cite{Webb1999} An indication of a varying $\alpha$ was detected by comparing metal absorptions at high redshift with corresponding transitions that were measured in the laboratory. As the observed transitions have in general a different dependence on $\alpha$, a variation manifests itself as a frequency shift of a certain line with respect to another. This is the basis of the Many-Multiplet-Method for probing a varying fine structure constant.\cite{Dzuba1999} The findings triggered numerous laboratory tests that compare transitions measured in different atoms and molecules over the course of a few years and thus probe a much shorter time scale for drifting constants.
In later work, Webb and co-workers found indication for a spatial variation of $\alpha$ in terms of a dipole across the Universe.\cite{Webb2011,King2012}

Spectroscopy provides a search ground for probing drifts in both $\alpha$ and $\mu$. While electronic transitions, including spin-orbit interactions, are sensitive to $\alpha$, vibrational, rotational and tunneling modes in molecules are sensitive to $\mu$. Hyperfine effects, such as in the Cs-atomic clock\cite{FlambaumTedesco2006,Berengut2011} and the 21-cm line of atomic hydrogen\cite{Tzanavaris2005}, depend on both $\alpha$ and $\mu$, as do $\Lambda$-doublet transitions in molecules\cite{Kozlov2009}.
The same holds for combined high-redshift observations of a rotational transition in CO and a fine structure transition in atomic carbon,\cite{Levshakov2012} placing a tight constraint on the variation of the combination $\alpha^2\mu$ at a redshift as high as $z=5.2$.
Within the framework of Grand Unification schemes theories have been developed that relate drifts in $\mu$ and $\alpha$ via

\begin{equation}
\frac{\Delta\mu}{\mu} = R \frac{\Delta\alpha}{\alpha}
\label{GUT}
\end{equation}

\noindent
where the proportionality constant $R$ should be large, on the order of $20-40$, even though its sign is not predicted.\cite{Calmet2002,Flambaum2004}
This would imply that $\mu$ is a more sensitive test ground than $\alpha$ when searching for varying constants.

The sensitivity of a spectroscopic experiment searching for a temporal variation of $\mu$ (and similarly for $\alpha$) can be expressed as
\begin{equation}
\left (\frac{\partial \mu}{\partial t} \right )\left / \mu = \left (\frac{\partial\nu}{\nu} \right ) \right / \left (K_\mu \Delta t\right ),
\label{eq:detect_variation_int}
\end{equation}
\noindent
assuming a linear drift. Here $({\partial{\mu}}/{\partial t})/{\mu}$ is the fractional rate of change of $\mu$, ${\partial \nu}/{\nu}$ is the fractional frequency precision of the measurement, $K_\mu$ is the inherent sensitivity of a transition to a variation of $\mu$, and $\Delta t$ is the time interval that is probed in the experiment. For a sensitive test, one needs transitions that are observed with a good signal to noise and narrow linewidth, and that exhibit high $K_{\mu}$. In order to detect a possible variation of $\mu$ at least two transitions possessing a different sensitivity are required.

Note that, for detecting a variation of $\mu$, it is not necessary to actually determine its value. In fact, in most cases this is impossible, since the exact relation between the value of $\mu$ and the observed molecular transitions is not known. Only for the most simple systems such as \ce{H2^+} and \ce{HD^+}, recently it became feasible to directly extract information on the value of $\mu$ from spectroscopic measurements~\cite{SchillerKorobov2005,KorobovZhong2012}. So far, the numerical value of the proton-electron mass ratio, $\mu =1836.152 672 45 \, (75)$, is known at a fractional accuracy of $4.1 \times 10^{-10}$ and included in CODATA\cite{Mohr2012,FootnoteA}, while constraints on the fractional change of $\mu$ are below $10^{-14}$/yr, as will be discussed in this paper.

The most stringent independent test of the time variation of $\mu$ in the current epoch was set by comparing vibrational transitions in \ce{SF6} with a cesium fountain over the course of two years. The \ce{SF6} transitions were measured with a fractional accuracy of $\sim$10$^{-14}$ and have a sensitivity of $K_\mu=-\tfrac{1}{2}$, whereas the sensitivity coefficient of the \ce{Cs} transition is $K_\mu\approx -1$\cite{FlambaumTedesco2006,Berengut2011}, resulting in a limit on the variation of $\Delta \mu/\mu$ of $5.6 \times 10^{-14}$/yr.\cite{Shelkovnikov2008}

In order to improve the constraints -- or to detect a time-variation -- attention has shifted to molecular species that possess transitions with greatly enhanced sensitivity coefficients. Unfortunately, the transitions that have an enhanced sensitivity are often rather exotic, i.e., transitions involving highly exited levels in complex molecules that pose considerable challenges to experimentalists and are difficult or impossible to observe in galaxies at high red-shift. Nevertheless, a number of promising systems have been identified that might lead to competitive laboratory and astrophysical tests in the near future.

In this Perspective we review the current status of laboratory and astrophysical tests on a possible time-variation of $\mu$. In particular we outline the procedures for determining the sensitivity coefficients for the different molecular species. Reviews on the topic of varying constants were presented by Uzan~\cite{Uzan2003}, approaching the subject from a perspective of fundamental physics, and by Kozlov and Levshakov~\cite{KozlovLevshakov2013}, approaching the topic from a molecular spectroscopy perspective.

\section{Definition of sensitivity coefficients}
The induced frequency shift of a certain transition as a result of a drifting constant is -- at least to first order -- proportional to the fractional change in $\alpha$ and $\mu$ and is characterized by its sensitivity coefficients $K_\alpha$ and $K_\mu$ via

\begin{equation}
\frac{\Delta\nu}{\nu} = K_{\alpha}\frac{\Delta\alpha}{\alpha}+K_{\mu}\frac{\Delta\mu}{\mu},
\label{eq:KlphaKmu}
\end{equation}

\noindent
where $\Delta\nu/\nu=(\nu_\text{obs}-\nu_0)/\nu_0$ is the fractional change in the frequency of the transition and $\Delta\mu/\mu=(\mu_\text{obs}-\mu_0)/\mu_0$ is the fractional change in $\mu$, both with respect to their current-day values. From Eq.~\eqref{eq:KlphaKmu} we can derive an expression for $K_\mu$ (and similarly for $K_\alpha$)

\begin{equation}
K_\mu = \frac{\mu}{E_e-E_g}\left (\frac{d E_e}{d\mu} -\frac{d E_g}{d\mu}\right ),
\label{eq:Kmu}
\end{equation}

\noindent
where $E_g$ and $E_e$ refer to the energy of the ground and excited state, respectively. Note that the concept of a ground state may be extended to any lower state in a transition, even if this corresponds to a metastable state or a short-lived excited state in a molecule. This definition of $K_\mu$ yields opposite signs to that used in Refs. [\!\!\citenum{Reinhold2006,Ubachs2007,Salumbides2012}].

Although electronic transitions in atoms are sensitive to $\alpha$, they are relatively immune to a variation of $\mu$. For instance, the frequency of the radiation emitted by a hydrogen-like element with nuclear charge $Ze$ and mass number $A$ in a transition between levels $a$ and $b$ is given by

\begin{equation}
\nu_{ab}=Z^2\frac{\mu_\text{red}}{m_e}R_\infty \left (\frac{1}{n_a^2}-\frac{1}{n_b^2} \right ),
\label{eq:Rydberg}
\end{equation}

\noindent
where $\mu_\text{red}={{A m_p}{m_e}}/(A m_p+m_e)$ and $R_\infty$ is the Rydberg constant. In order to find the sensitivity coefficients of these transitions we apply Eq.~\eqref{eq:Kmu} and obtain

\begin{equation}
K_\mu = \frac{1}{1+A\mu},
\label{eq:RydbergKmu}
\end{equation}

\noindent
resulting in sensitivity coefficients of $5.4\times 10^{-4}$ for the transitions of the Lyman series in atomic hydrogen ($A=1$).

Let us now turn to transitions in molecules. Within the framework of the Born-Oppenheimer approximation, the total energy of a molecule is given by a sum of uncoupled energy contributions, hence, we may rewrite Eq.~\eqref{eq:Kmu} as

\begin{equation}
K_\mu \approx \frac{\sum\nolimits_i K_\mu^i\Delta E_i}{\sum\nolimits_i \Delta E_i},
\label{eq:toy}
\end{equation}

\noindent
where the summation index $i$ runs over the different energy contributions, such as electronic, vibrational, and rotational energy. It is generally assumed that the neutron-to-electron mass ratio follows the same behavior as the proton-to-electron mass ratio and no effects depending on quark structure persist\cite{Dent2007}. Under this assumption all baryonic matter may be treated equally and $\mu$ is proportional to the mass of the molecule. Hence, from the well-known isotopic scaling relations we find $K_\mu^\text{el}=0$, $K_\mu^\text{vib}=-\tfrac{1}{2}$, and $K_\mu^\text{rot}=-1$.


The inverse dependence of the sensitivity coefficient on the transition frequency suggests that $K_\mu$ is enhanced for near-degenerate transitions, i.e., when the different energy contributions in the denominator of Eq.~\eqref{eq:toy} cancel each other. This enhancement is proportional to the energy that is being cancelled and to the difference in the sensitivity coefficients of the energy terms. Since in general $E_\text{el}\gg E_\text{vib}\gg E_\text{rot}$, cancellations between electronic, vibrational and rotational energies are unexpected. Nevertheless, transitions with enhanced sensitivity due to a cancellation of vibrational and electronic energies have been identified in \ce{Cs2}~\cite{DeMille2008} and \ce{NH+}~\cite{Beloy2011}. Whereas cancellations between electronic and vibrational energies are purely coincidental, near-degeneracies occur as a rule in more complex molecules such as molecular radicals or poly-atomic molecules. These molecules have additional energy contributions that are comparable in magnitude to rotational and vibrational energies and exhibit a different functional dependence on $\mu$. For instance, molecules in electronic states with non-zero electronic angular momentum have fine-structure splittings that are comparable to vibrational splittings in heavy molecules\cite{FlambaumKozlov2007} and to rotational splittings in light molecules~\cite{BethlemUbachs2009}. Likewise, molecules that possess nuclear spin have hyperfine splittings that can be comparable to rotational splittings\cite{Flambaum2006}. In polyatomic molecules, splittings due to classically-forbidden large-amplitude motions, such as inversion~\cite{vanVeldhoven2004,FlambaumKozlov2007NH3,KozlovLevshakov2011} or internal rotation\cite{Jansen2011PRL,KozlovLevshakov2013}, can be comparable to rotational splittings. Finally, the Renner-Teller splitting, that originates from the interaction between electronic and vibrational angular momenta in linear polyatomic molecules, can be comparable to rovibrational splittings\cite{Kozlov2013}. \\

As discussed in the introduction, the sensitivity of a test depends both on the sensitivity coefficient and the fractional precision of the measured transition (see Eq.~(\ref{eq:detect_variation_int}). For enhancements originating from cancellations  between different modes of energy the  sensitivity scales as the inverse frequency – i.e., when two energy terms in the numerator of Eq.~(\ref{eq:Kmu}) are very similar the sensitivity coefficient becomes large while the transition frequency becomes small. The resolution of astrophysical observations are usually limited by Doppler broadening which implies that the fractional precision, $\delta \nu/\nu$, is independent of the frequency.  Thus, for astrophysical tests the advantage of low frequency transitions with enhanced sensitivity is evident. For laboratory tests, the motivation for choosing low frequency transitions is less obvious. Due to the advances in frequency comb and optical clock techniques, the fractional precision of optical transitions has become superior to those in the microwave domain.\cite{Chou2010,Nicholson2012} It was therefore argued by Zelevinsky~\emph{et al.}\cite{Zelevinsky2008} and others, that the best strategy for testing the time-variation of fundamental constants is to measure an as large as possible energy interval and accept the rather limited sensitivity coefficient that is associated with it.
It may be true that optical clocks have a better fractional accuracy but microwave measurements still have a smaller absolute uncertainty. For instance, the most accurate optical clock based on a transition in Al$^+$ at 267~nm (1.12~PHz) has a fractional accuracy of $2.3 \times 10^{-17}$, which corresponds to an absolute uncertainty of 27~mHz~\cite{Rosenband2008}, while the most accurate microwave clock, based on a transition in Cesium at 9.2~GHz has a fractional accuracy of $2 \times 10^{-16}$ corresponding to an absolute uncertainty of 2~$\mu$Hz~\cite{Bize2005}. It thus make sense to measure transitions in the microwave region, but only if favorable enhancement schemes are available. An additional advantage is that, in some well-chosen cases, transitions with opposite sensitivity coefficients can be used to eliminate systematic effects.

The remainder of this paper can be divided into two parts. In the first part, consisting of Secs.~\ref{sec:hydrogen} and~\ref{sec:radicals}, the use of diatomic molecules in studies of a time-varying $\mu$ is discussed. In particular, Sec.~\ref{sec:hydrogen} reviews the calculation of sensitivity coefficients for rovibronic transitions in molecular hydrogen and carbon monoxide and describes how these transitions are used to constrain temporal variation of $\mu$ on a cosmological time scale. Section~\ref{sec:radicals} shows that the different mass dependence of rotational and spin-orbit constants results in `accidental' degeneracies for specific transitions. The second part of the paper consists of Secs.~\ref{sec:inversion} to~\ref{sec:internal_rotation} and discusses the use of polyatomic molecules, in particular those that possess a classically-forbidden tunneling motion.

\section{Testing the time independence of $\mu$ using diatomic molecules}
\label{sec:diatomics}

\subsection{Transitions in molecular hydrogen and carbon monoxide\label{sec:hydrogen}}
Molecular hydrogen has been the target species of choice for $\mu$ variation searches on a cosmological time scale, in particular at higher redshifts ($z > 2$). The wavelengths of the Lyman and Werner absorption lines in \ce{H2} and \ce{HD} can be detected in high-redshifted interstellar clouds and galaxies in the line of sight of quasars and may be compared with accurate measurements of the same transitions performed in laboratories on earth. While Thompson proposed using high-redshift \ce{H2} lines as a search ground for a varying proton-electron mass ratio~\cite{Thompson1975}, Varshalovich and Levshakov first calculated $K_{\mu}$ sensitivity coefficients for the \ce{H2} molecule\cite{Varshalovich1993}. Later updated values for sensitivity coefficients of \ce{H2} were obtained in a semi-empirical fashion, based on newly established spectroscopic data~\cite{Reinhold2006,Ubachs2007}, and via \emph{ab initio} calculations.\cite{Meshkov2006}

In the semi-empirical approach, rovibrational level energies of the relevant electronic states are fitted to a Dunham expansion\cite{Dunham1932}

\begin{equation}
E(\nu,J)= \sum_{k,l}Y_{kl}\left (\nu +\tfrac{1}{2} \right )^k \left [J\left (J+1 \right ) - \Lambda^2 \right ]^l,
\label{eq:Dunhamex}
\end{equation}

\noindent
where $\Lambda$ is the projection of the orbital angular momentum on the molecular axis, i.e., $\Lambda = 0$ and $1$ for $\Sigma$ and $\Pi$ states, respectively, and $Y_{kl}$ are the fitting parameters. The advantage of the Dunham representation of molecular states is that the coefficients scale to first order as $Y_{kl}\propto \mu_\text{red}^{l-k/2}$, with $\mu_\text{red}$ the reduced mass of the molecule\cite{Dunham1932,Ubachs2007}. The coefficients from the Dunham expansion can thus be used to determine the sensitivity coefficients through

\begin{multline}
\frac{d E}{d \mu} = \sum_{k,l}\frac{d Y_{kl}}{d\mu}\left (\nu +\tfrac{1}{2} \right )^k \left [J\left (J+1 \right ) - \Lambda^2 \right ]^l,\\
\text{with } \frac{d Y_{kl}}{d\mu}\approx -\frac{Y_{kl}}{\mu} \left (l+\frac{k}{2} \right ).
\label{eq:Danhamder}
\end{multline}

By inserting Eqs.~\eqref{eq:Dunhamex} and~\eqref{eq:Danhamder} into Eq.~\eqref{eq:Kmu}, sensitivity coefficients are obtained within the Born-Oppenheimer approximation. The mass dependence of the potential minima of ground and excited states is partly accounted for by including the adiabatic correction. Neglecting the dependence on the nuclear potential, its effect is approximated to that of the normal mass shift or Bohr shift\cite{Bohr1913}, $R_\text{H}/R_\infty=m_p/(m_p+m_e)$, on the levels of an electron bound to an \ce{H2+} core, due to the finite mass of the latter

\begin{equation}
\Delta E_\text{ad} = -\frac{\Delta E_\infty}{2(\mu+1)}=-\frac{\Delta E(\mu)}{2\mu+1},
\label{eq:DEad}
\end{equation}

\noindent
where $\Delta E_\mu$ is the difference of the empirical $Y_{00}$ values of the (deperturbed) $B ^1\Sigma^+_u$ or $C ^1\Pi_u$ state and the $X ^1\Sigma^+_g$ ground state. The mass dependence of Eq.~\eqref{eq:DEad} introduces an additional term that should be included in the parenthesis of Eq.~\eqref{eq:Kmu} representing the adiabatic correction

\begin{equation}
\frac{d}{d\mu}\Delta E_\text{ad} = -\frac{\Delta E_\text{ad}}{\mu+1}.
\label{eq:DEadder}
\end{equation}

In order to account for nonadiabatic interaction, mixing between different electronic states should be included. In Refs.~[\!\!\citenum{Reinhold2006,Ubachs2007}] a model is adopted in which the multi-dimensional problem is approximated by incorporating only the interaction of the dominant electronic states. The values for the resulting interaction matrix elements are obtained from a fit to the experimental data. This procedure provides both the deperturbed level energies to which the Dunham coefficients are fitted, as well as the superposition coefficients of the mixed states, $c_i$. The sensitivity coefficients for the perturbed states are given by

\begin{equation}
K_\mu=\sum_i c_i^2 K_\mu^i,
\label{eq:Kmuperturbed}
\end{equation}

\noindent
where $i=0$ refers to the state under consideration and $K_\mu^i$ are the sensitivity coefficients of the perturbing states.
In particular for some levels where a strong interaction between $B\,^1\Sigma_u^+$ and $C\,^1\Pi_u$ states occurs the non-adiabatic interaction contributes significantly to the values of $K_{\mu}$.

\begin{figure}[bt]
\centering
\includegraphics[width=1\columnwidth]{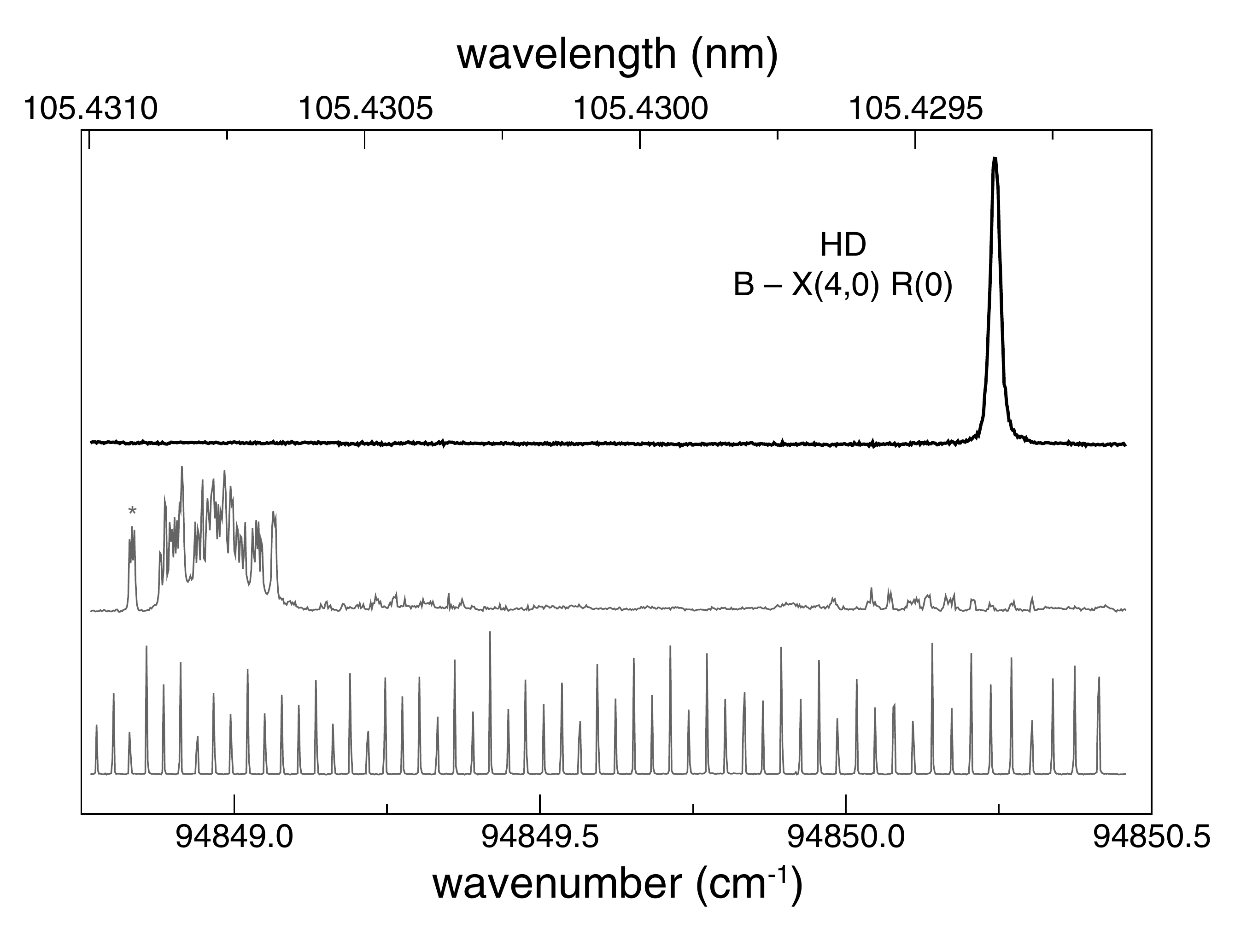}
\caption{Representative high-resolution spectrum. Recording of the $R(0)$ line in the $B-X(4,0)$ band of \ce{HD} (upper) with etalon markers (lower) and an \ce{I2}-saturation spectrum (middle) for calibration. The line marked with an asterisk (*) is the $\mathrm{a}_2$ hyperfine component of the $B-X(8,4)\,R(49)$ transition line in \ce{I2} at $15\,808.13518$\,\wn used as an absolute reference~\cite{Xu2000}. Note that the \ce{I2} and etalon spectra are taken at the fundamental, whereas the XUV axis shown is the 6th harmonic.
\label{fig:H2-spec-line}}
\end{figure}

The procedures, following this semi-empirical (SE) procedure outlined in the above, yield $K_{\mu}$ coefficients for the Lyman lines (in the $B\,^1\Sigma_u^+$ - $X\,^1\Sigma_g^+$ system) and Werner lines (in the $C\,^1\Pi_u$ - $X\,^1\Sigma_g^+$ system) in the range (-0.05, +0.02). These results agree with values obtained from \emph{ab initio} calculations (AI) within $\Delta K_{\mu} = K_{\mu}^{AI} - K_{\mu}^{SE} < 3 \times 10^{-4}$, so at the 1\% level, providing confidence that a reliable set of sensitivity coefficients for H$_2$ is available. For the HD molecule a set of $K_{\mu}$ coefficients was obtained via \emph{ab initio} calculations.\cite{Ivanov2010}\\

A full set of accurate laboratory wavelengths was obtained in spectroscopic studies with the Amsterdam narrowband extreme ultraviolet (XUV) laser setup. Coherent and tunable radiation at wavelengths $92-112$\,nm is produced starting from a Nd:VO$_4$-pumped continuous wave (CW) ring dye laser, subsequent pulse amplification in a three-stage traveling-wave pulsed dye amplifier, frequency doubling in a KDP-crystal to produce UV-light, and third harmonic generation in a pulsed jet of Xe gas~\cite{Ubachs1997}. The spectroscopy of the strong dipole allowed transitions in the Lyman bands and Werner bands was performed in a configuration with a collimated beam of \ce{H2} molecules perpendicularly crossing the overlapping XUV and UV beams via the method of $1+1$ resonance-enhanced photo-ionization. Calibration of the absolute frequency scale in the XUV was established via comparison of the CW-output of the ring laser with on-line recording of saturated absorption lines of \ce{I2} and fringes of a Fabry-Perot interferometer, which was stabilized against of \ce{HeNe} laser. Wavelength uncertainties, for the major part related to residual Doppler effects, AC-Stark induced effects and frequency chirp in the pulsed dye amplifier, as well as to statistical effects, were carefully addressed leading to calibrated transition frequencies of the Lyman and Werner band lines in the range $92-112$\,nm at an absolute accuracy of $0.004$\,\wn\ or $0.000004$\,nm, corresponding to a relative accuracy of $5 \times 10^{-8}$.  A detailed description of the experimental procedures and of the results is given in a sequence of papers~\cite{Philip2004a,Philip2004b,Ivanov2008b}. Similar investigations of the XUV-laser spectrum of HD were performed in view of the fact that HD lines were also observed in high-redshift spectra towards quasar sources~\cite{Hollenstein2006,Ivanov2008}.

Additional spectroscopic studies of \ce{H2} were performed assessing the level energies in these excited states in an indirect manner, thereby verifying and even improving the transition frequencies in the Lyman and Werner bands~\cite{Salumbides2008,Bailly2010}. The data set of laboratory wavelengths obtained for both \ce{H2} and \ce{HD}, has reached an accuracy that can be considered exact for the purpose of comparison with quasar data, where accuracies are never better than $10^{-7}$.  A typical recording of an HD lines is shown in Fig.~\ref{fig:H2-spec-line}. A full listing of all relevant parameters on the laboratory absorption spectrum of \ce{H2} and \ce{HD}, including information on the intensities, is made available in digital form in the supplementary material of Ref.~[\onlinecite{Ivanov2008}].

\begin{figure}[bt]
\centering
\includegraphics[width=1\columnwidth]{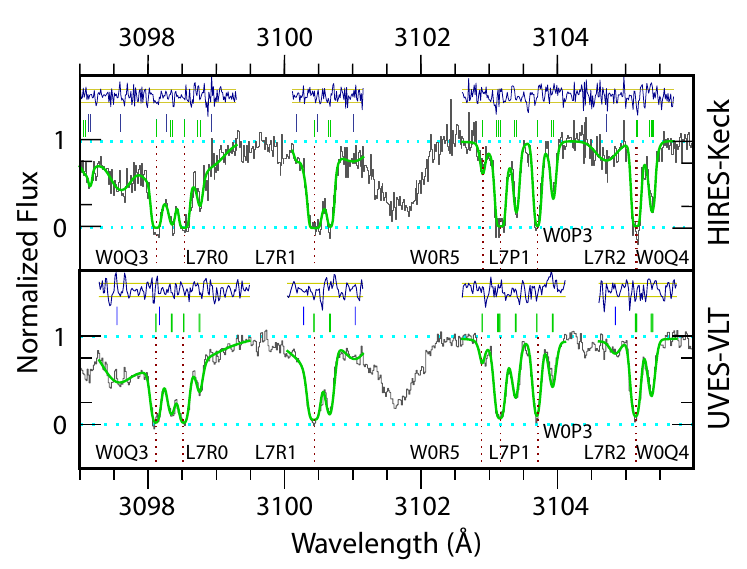}
\caption{Comparison between the spectrum of Q2123-005 in the $3097-3106$\,\AA\ range observed with HIRES-Keck\cite{Malec2010} (upper panel) and UVES-VLT\cite{vanWeerdenburg2011} (lower panel). For both panels, fits to the molecular hydrogen lines are shown as the solid green lines and their velocity components are indicated by the tick marks that are shown above the spectrum. Tick marks indicating the positions of Lyman-$\alpha$ lines and \ce{Fe\,II} lines are shown with a slight offset. \ce{H2} spectral line identifications are shown at the bottom. Residuals from the fit are shown above the observed spectra.
\label{fig:keck_vlt}}
\end{figure}

High quality data on high redshift absorbing systems, in terms of signal-to-noise (S/N) and resolution, is available only for a limited number of objects. In view of the transparency window of the earth's atmosphere ($\lambda > 300$\,nm) absorbing systems at $z>2$ will reveal a sufficient number of lines to perform a $\Delta\mu/\mu$ constraining analysis. The systems observed and analyzed so far are: Q0347-383 at $z_{\rm{abs}}= 3.02$, Q0405-443 at $z_{\rm{abs}}= 2.59$, Q0528-250 at $z_{\rm{abs}}= 2.81$, Q2123-005 at $z_{\rm{abs}}= 2.05$, and Q2348-011 at $z_{\rm{abs}}= 2.42$. Note that the objects denoted by "Q" are background quasars, which in most studies focusing on H$_2$ spectra are considered as background light sources, and are indicated by their approximate right ascension (in hours, minutes and seconds) as a first coordinate and by their declination (in degrees, arcminutes and arcseconds, north with "+" and south with "-") as a second coordinate. Hence Q0347-383 refers to a bright quasar located at RA =03:49:43.64 and dec =-38:10:30.6 in so-called J2000 coordinates (the slight discrepancies in numbers relate to the fact that most quasars were discovered some 30 years ago, in the epoch when the B1950 coordinate system was in use; hence they derive their names from the older, shifted coordinate frame). These coordinates imply that Q0347-383 is observable during night-time observations in October and a few months before and after. This quasar source is known to be located at $z_{\rm{emis}}=3.21$ from a Lyman-$\alpha$ intensity peak in its emission spectrum, while the absorbing galaxy containing one or more clouds with H$_2$ is at $z_{\rm{abs}}=3.02$. From the analysis of the redshifted H$_2$ spectrum a 7-digit accuracy value for the redshift is obtained, in the case of Q0347-383 $z_{\rm{abs}}=3.024\,899\,0 (12)$ \cite{Reinhold2006}. Such an accurate determination of $z_{\rm{abs}}$ is required for the $\mu$-variation analysis, since it sets the exact value of the Doppler shift of the absorbing cloud.

Relevant parameters for the analysis are the \ce{H2} column density, which should be sufficient to yield absorption of at least the lowest $J$-levels, hence $N$(\ce{H2}) $> 10^{14}$\,cm$^{-2}$ and lower than $10^{19}$\,cm$^{-2}$ to avoid full saturation of the lines, and the brightness of the background quasar which should produce a high S/N in a reasonable amount of observing time. The absorbing system toward Q2123-005 has the favorable condition that the magnitude of the quasar background source ($R_\text{mag}= 15.8$) is the brightest of all \ce{H2} bearing systems observed so far. This system has been observed from both the Very Large Telescope (Paranal, Chile), equipped with the Ultraviolet-Visible Echelle Spectrometer (UVES) and with the Keck Telescope (Hawaii, USA) equipped with the HIRES spectrometer.
For a comparison of observed spectra see Fig.~\ref{fig:keck_vlt}. The results from the analyses are $\Delta\mu/\mu = (5.6 \pm 5.5_{stat} \pm 2.9_{syst}) \times 10^{-6}$ for the Keck spectrum\cite{Malec2010} and $\Delta\mu/\mu = (8.5 \pm 3.6_{stat} \pm 2.2_{syst}) \times 10^{-6}$ for the VLT spectrum\cite{vanWeerdenburg2011}, are tightly constraining and in good agreement with each other. This result eases concerns on systematic effects associated with each of the instruments. Brightness of the other background quasars is typically $R_\text{mag}=17.5$, while Q2348-011 is the weakest with $R_\text{mag}=18.3$. The latter only delivered a poor constraint for reasons of low brightness and from a second damped-Lyman absorber taking away many \ce{H2} lines by its Lyman cutoff~\cite{Bagdonaite2012}.

\begin{figure*}
\centering
\includegraphics[width=0.95\textwidth]{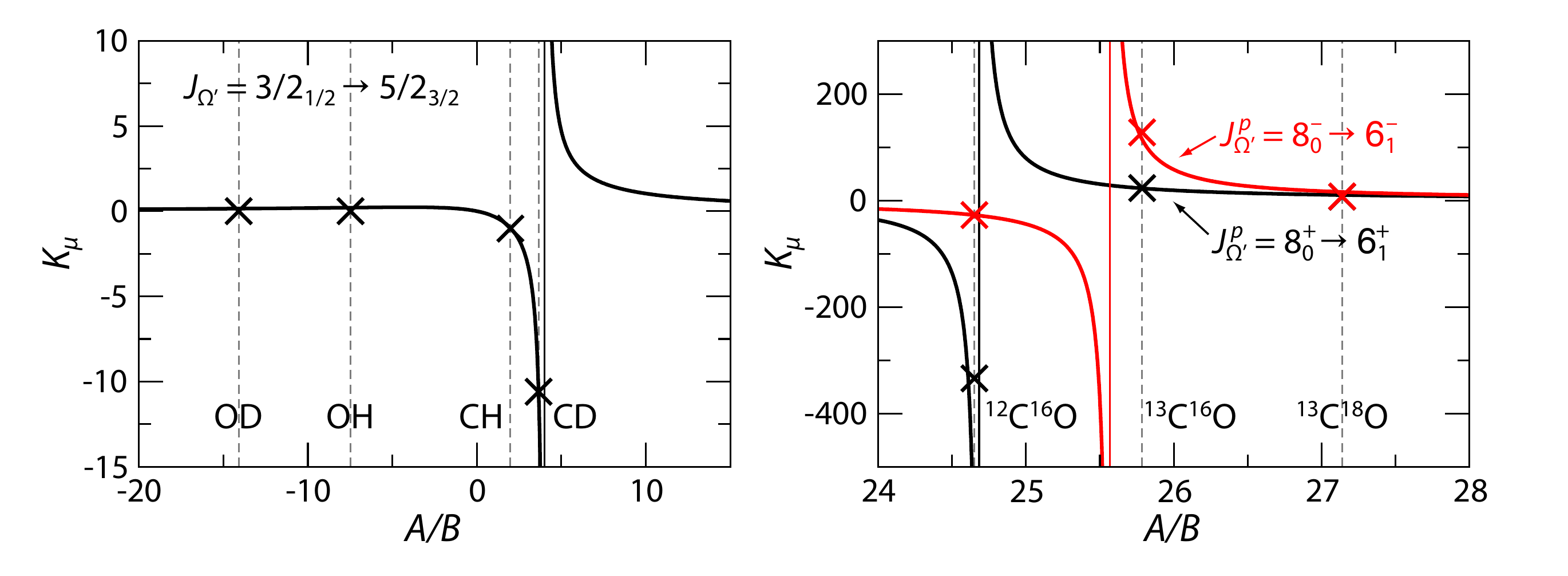}
\caption{Sensitivity coefficients $K_\mu$ as a function of $A/B$ for $J_{{\Omega}\prime}=3/2_{1/2}\rightarrow 5/2_{3/2}$ transitions in diatomic molecules in a $^2\Pi$ electronic state, calculated using Eq.~\eqref{eq:SOrotHmatrix} (left panel, after de Nijs \emph{et al.}\cite{DeNijs2012}) and $J_{\Omega\prime}^p=8_0^\pm\rightarrow 6_1^\pm$ transitions in the $a ^3\Pi$ state of several \ce{CO} isotopologues (right panel). The crosses indicate sensitivity coefficients that were calculated while taking higher-order contributions to the Hamiltonian into account\cite{DeNijs2011,DeNijs2012}.
\label{fig:CHandCO}}
\end{figure*}

Since the number of suitable \ce{H2} absorber systems at high redshift is rather limited, additional schemes are required to improve the current constraint on $\mu$ variation at redshifts $z>1$. Recent observations of vacuum ultraviolet transitions in carbon monoxide at high redshift\cite{Srianand2008,Noterdaeme2009,Noterdaeme2010,Noterdaeme2011} make \ce{CO} a promising target species for probing variation of $\mu$. An additional advantage of the \ce{CO} $A-X$ bands is that its wavelengths range from $130-154$\,nm, that is, at lower wavelengths than Lyman-$\alpha$, so that the \ce{CO} spectral features in typical quasar spectra will fall outside the region of the so-called Lyman-$\alpha$ forest (provided that the emission redshift of the quasar $z_\text{em}$ is not too far from the redshift $z_\text{abs}$ of the intervening galaxy exhibiting the molecular absorption). The occurrence of the Lyman-forest lines is a major obstacle in the search for $\mu$ variation via molecular hydrogen lines.

In order to prepare for a $\mu$-variation analysis, accurate laboratory measurements on the $A-X$ system of CO were performed, using laser-based excitation and Fourier-transform absorption spectroscopy~\cite{Salumbides2012}, yielding transition frequencies at an accuracy better than $\Delta\lambda/\lambda = 3 \times 10^{-7}$. Also a calculation of $K_{\mu}$ sensitivity coefficients was performed, which required a detailed analysis of the structure of the $A^1\Pi$ state of CO and its perturbation by a number of nearby lying singlet and triplet states.\cite{Niu2013}

\subsection{Near-degeneracies in diatomic radicals\label{sec:radicals}}

In the previous section we discussed sensitivity coefficients for transitions in diatomics with closed-shell electronic states, that is, molecules that have zero electronic orbital angular momentum.

Let us now turn to diatomic open-shell molecules in a $^2\Pi$ electronic state that have a nonzero projection of orbital angular momentum along the molecular axis. The overall angular momentum $\mathbf{J}$ depends on the coupling between the orbital angular momentum $\mathbf{L}$, the spin angular momentum $\mathbf{S}$, and the rotational angular momentum $\mathbf{R}$. Depending on the energy scales that are associated with these momenta, the coupling between the vectors is described by the different Hund's cases.

When only rotation and spin-orbit coupling are considered, the Hamiltonian matrix for a $^2\Pi$ electronic state in a Hund's case (a) basis is given by\cite{BrownCarrington}

\begin{multline}
\begin{pmatrix}
\tfrac{1}{2}A+Bz		& -B\sqrt{z} \\
-B\sqrt{z}				& -\tfrac{1}{2}A+B\left (z+2 \right )
\end{pmatrix},\\\text{with }z=\left (J+\tfrac{1}{2} \right )^2 -1
\label{eq:SOrotHmatrix}
\end{multline}

\noindent
where $A$ and $B$ refer to the spin-orbit and rotational constant, respectively. For a given value of $J$, the lower energy level is labelled as $F_1$ and the upper as $F_2$. The eigenfunctions of the Hamiltonian matrix \eqref{eq:SOrotHmatrix} are

\begin{equation}
\ket{F_2}=a_J\ket{\tfrac{3}{2}}-b_J\ket{\tfrac{1}{2}}\text{ and }
\ket{F_1}=b_J\ket{\tfrac{3}{2}}+a_J\ket{\tfrac{1}{2}},
\end{equation}

\noindent
where
\begin{equation}
a_J^2=\frac{X+(A-2B)}{2X},\text{ and }
b_J^2=\frac{X-(A-2B)}{2X},
\end{equation}

\noindent
and
\begin{equation}
X=\sqrt{(A-2B)^2+4B^2z}.
\end{equation}

It is instructive to analyze the sensitivity coefficients of transitions within these molecules as a function of $A/B$. These transitions can be divided into two categories; transitions within a spin-orbit manifold and transitions between adjacent spin-orbit manifolds. In the limit of large $|A/B|$, transitions within a $\Omega$ manifold become purely rotational having $K_\mu=-1$, while transitions between different $\Omega$ manifolds, become purely electronic, and therefore have $K_\mu=0$. When $A\sim Bz$, the spin-orbit manifolds become mixed and the sensitivity of the different types of transitions lies between 0 and -1\cite{DeNijs2012}. Three distinct situations, illustrated for a single transition in the left-hand side of Fig.~\ref{fig:CHandCO}, can be identified; (i) When $A=0$ all transitions have a sensitivity coefficient of $-1$. (ii) When $A=2B$, $a_J=b_J=1/\sqrt{2}$ and the spin-orbit manifolds are completely mixed. This also results in sensitivity coefficients of $K_\mu=-1$. (iii) Finally, when $A=4B$, the levels $F_1(J)$ and $F_2(J-1)$ are degenerate for each value of $J$. This case (b) `behavior' (zero spin-orbit splitting) gives rise to an enhancement of the sensitivity coefficient for transitions that connect these two states. However, it was shown by de Nijs~\emph{et al.}\cite{DeNijs2012} that the same conditions that led to the enhancement of the sensitivity coefficients also suppress the transition strength, leading them to conclude that one-photon transitions between different spin-orbit manifolds of molecular radicals are either insensitive to a variation of $\mu$ or too weak to be of relevance in astrophysical searches for variation of $\mu$.

\begin{figure}[tb]
\centering
\includegraphics[width=1\columnwidth]{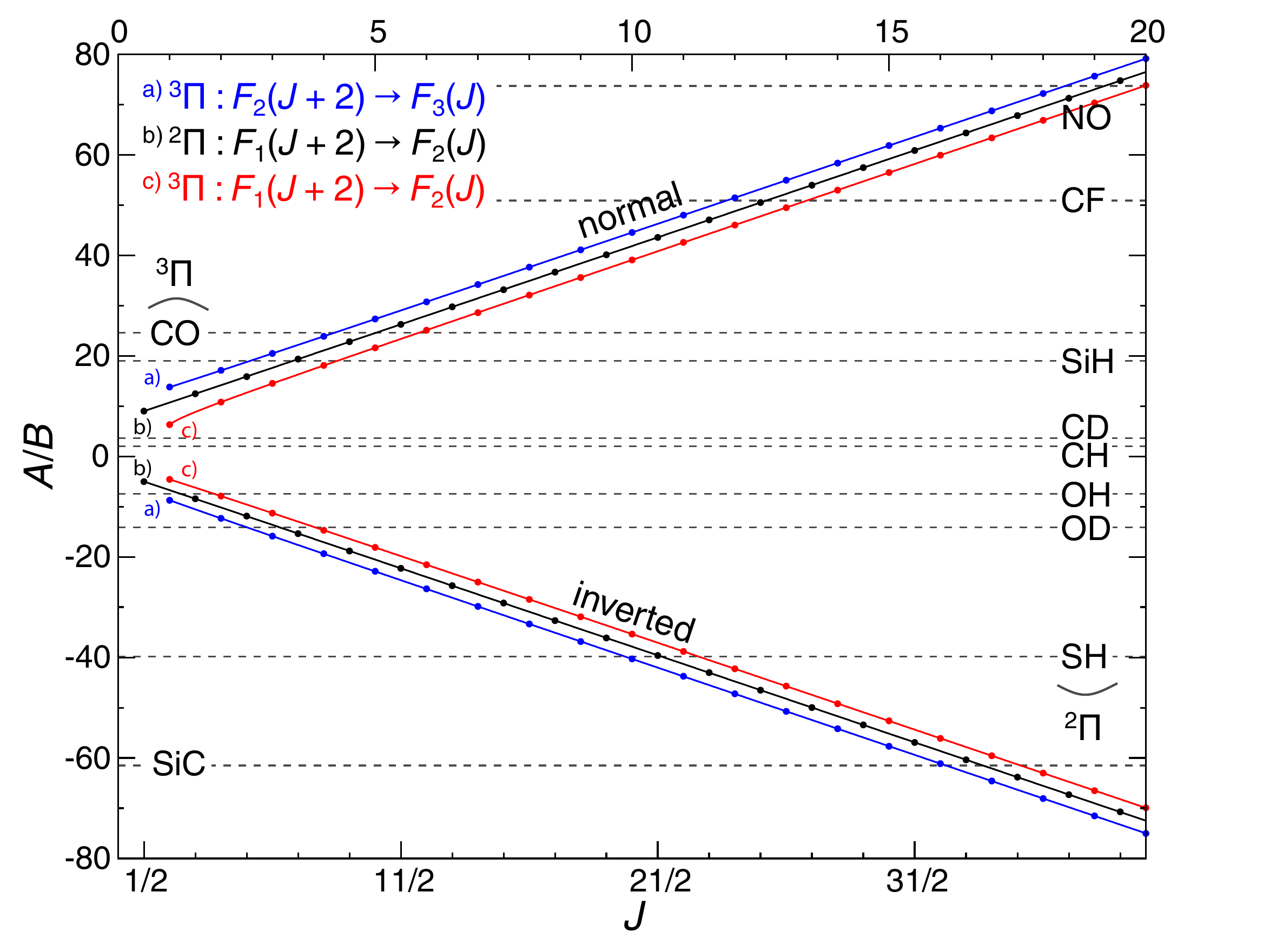}
\caption{Relation between $A/B$ and the value of $J$ at which a near degeneracy occurs for diatomic molecules in doublet and triplet $\Pi$ states. The curves were calculated using a simplified model that neglects lambda and hyper-fine splitting.
\label{fig:2Pi3Pi_res}}
\end{figure}

This problem disappears when two-photon transitions are considered, as was done by Bethlem and Ubachs\cite{BethlemUbachs2009} for \ce{CO} in its metastable $a\,^3\Pi$ state, which is perhaps
the best studied excited triplet system of any molecule.\cite{Freund1965,Wicke1972,Saykally1987,Yamamoto1988}
On the right-hand side of Fig.~\ref{fig:CHandCO}, sensitivity coefficients for the $J_\Omega^p = 6_1^\pm - 8_0^\pm$ transitions in \ce{CO} are shown as a function of $A/B$, calculated using a simplified Hamiltonian matrix for the $a\,^3\Pi$ state. Note that "+" and "-" signs refer to $\Lambda$-doublet components of opposite parity. Crosses, also shown in the figure, indicate sensitivity coefficients that were calculated using a full molecular Hamiltonian.\cite{DeNijs2011} From the figure it can be seen that resonances occur near $A/B\sim 25$ which is close to the $A/B$ values for the \ce{^{12}C^{16}O} and \ce{^{13}C^{16}O} isotopologues. When combined, the $6_1^+\rightarrow 8_0^+$ transition in \ce{^{12}C^{16}O} and the $8_0^-\rightarrow 6_1^-$ transition in \ce{^{13}C^{16}O} have a sensitivity that is almost 500 times that of a pure rotational transition. An experiment to measure these transitions in a double-resonance molecular beam machine using a two-photon microwave absorption is currently under construction in our laboratory.\cite{BethlemUbachs2009}

The relation between $A/B$ and the value of $J$ at which a resonance is expected for two-photon transitions in diatomic molecules in doublet and triplet $\Pi$ states is shown in Fig.~\ref{fig:2Pi3Pi_res}. From this figure it is easily seen that no such resonances occur in \ce{CH} and \ce{CD}, because for these molecules the value of $A/B$ results in a fine-structure splitting that is smaller than the rotational splitting. Most other molecules have resonances that occur only for relatively high values of $J$, making these systems difficult to access experimentally. For molecules with $^2\Pi$ electronic states, we see that only \ce{OH}, \ce{OD}, and \ce{SiH} have near degeneracies for $J<10$, whereas \ce{CO} is the only molecule in a $^3\Pi$ electronic state with a resonance at low $J$.\\

In the present discussion only rotational transitions between different spin-orbit manifolds were considered. Darling first suggested that $\Lambda$-doublet transitions in OH could serve as a probe for a time-variation of $\alpha$
and $\mu$.\cite{Darling2003} These transitions were measured at high accuracy in a Stark-decelerated molecular beam by Hudson \emph{et al.}~\cite{Hudson2006} It was shown by Kozlov\cite{Kozlov2009} that $\Lambda$-doublet transitions in particular rotational levels of OH and CH have an enhanced sensitivity for $\mu$-variation, as a result of an inversion of the $\Lambda$-doublet ordering. For OH the largest enhancement occurs in the $J=9/2$ of the $\Omega=3/2$ manifold which lies 220~cm$^{-1}$ above the ground-state and gives rise to $K_{\mu} \sim 10^{3}$. For CH the largest enhancements occur in the $J=3/2$ of the $\Omega=3/2$, which lies only 18~cm$^{-1}$ above the ground state, however, the enhancement is on the order of 10.
Recently, Truppe \emph{et al.}\cite{Truppe2013} used Ramsey's separated zone oscillatory field technique to measure the 3.3 and 0.7\,GHz $\Lambda$-doublet transitions in \ce{CH} with relative accuracies of $9\times 10^{-10}$ and $3\times 10^{-8}$, respectively. By comparing their line positions with astronomical observations of \ce{CH} (and \ce{OH}) from sources in the local galaxy, they were able to constrain $\mu$-dependence on matter density effects (chameleon scenario) at $\Delta\mu/\mu<2.2\times 10^{-7}$.



\section{Large amplitude motion in polyatomic molecules}
\subsection{Tunneling inversion \label{sec:inversion}}
\begin{figure}
\centering
\includegraphics[width=1\columnwidth]{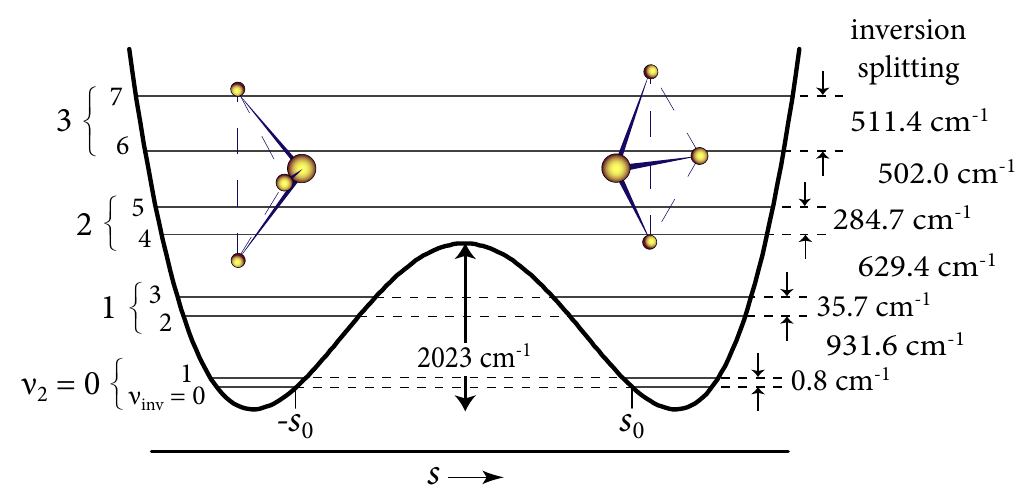}
\caption{Potential energy curve and lowest vibrational energy levels for the electronic ground state of \ce{NH3} as a function of the distance between the nitrogen atom and the plane of the hydrogen atoms, $s$. The classical turning points for the ground vibrational state, $\pm s_0$, are indicated as well. Due to tunneling through the potential barrier each vibrational level is split in a symmetric and antisymmetric component.
\label{fig:ammonia_potential}}
\end{figure}

In its electronic ground state, the ammonia molecule has the form of a regular pyramid, whose apex is formed by the nitrogen atom, while the base consists of an equilateral triangle formed by the three hydrogen atoms. Classically, the lowest vibrational states possess insufficient energy to allow the nitrogen atom to be found in the plane of the hydrogen atoms, as can be seen from the potential energy curve in Fig.~\ref{fig:ammonia_potential}. If the barrier between the two potential wells were of infinite height, the two wells would be totally disconnected and each energy eigenvalue of the system would be doubly degenerate. However, as the barrier is finite, quantum-mechanical tunneling of the nitrogen atom through the plane of the hydrogen atoms couples the two wells. This tunneling motion lifts the degeneracy, and the energy levels are split into doublets. The tunneling through the barrier with a height of 2023\,\wn is responsible for an energy splitting of 0.8\,\wn and 36\,\wn in the ground vibrational and first excited vibrational states, respectively. These energies are much smaller than the energy corresponding to the normal vibrational motion in a single well ($\tilde{\nu}_0=950$\,\wn), since the inversion of the molecule is severely hindered by the presence of the potential barrier.

An analytical expression for the inversion frequency has been calculated by Dennison and Uhlenbeck~\cite{Dennison1932}, who used the Wentzel-Kramers-Brillouin approximation to obtain

\begin{multline}
\omega_\mathrm{inv}=\frac{\omega_0}{\pi}e^{-G},\text{ with}\\\quad G=\frac{1}{\hbar}\int_{-s_0}^{s_0}\left [2\mu_{\text{red}} \left ( U(z) - E\right ) \right ]^{\tfrac{1}{2}}ds,
\label{eq:WKBsplitting}
\end{multline}
\noindent
with $\omega_0$ the energy of the vibration in one of the potential minima and $E$ the total vibrational energy.

Townes and Schawlow already noted that ``if the reduced mass is increased by a factor of 2, such as would be roughly done by changing from \ce{NH3} to \ce{ND3}, $\nu_\text{inv}$ decreases by $e^{6\left ( \sqrt{2}-1\right )}$ or a factor of 11.''\cite{TownesSchawlow1975}. Van Veldhoven \emph{et al.}\cite{vanVeldhoven2004} and Flambaum and Kozlov\cite{FlambaumKozlov2007NH3} pointed out that the strong dependence of the inversion splitting on the reduced mass of the ammonia molecule can be exploited to probe a variation of $\mu$.

To a first approximation the Gamow factor, $G$, is proportional to $\mu_\text{red}^{1/2}$ and the $\mu$ dependence of Eq.~\eqref{eq:WKBsplitting} can be expressed through

\begin{equation}
\nu_\text{inv}=\frac{a_0}{\sqrt{\mu_\text{red}}}e^{-a_1\sqrt{\mu_\text{red}}},
\label{eq:invfit}
\end{equation}

\noindent
where $a_0$ and $a_1$ are fitting constants. The sensitivity coefficient for the inversion frequency is thus given by

\begin{equation}
K_\mu^\text{inv}=-\tfrac{1}{2}a_1\sqrt{\mu_\text{red}}-\tfrac{1}{2}.
\label{eq:Kmuinvfit}
\end{equation}

\noindent
From a fit through the inversion frequencies of the different isotopologues of ammonia we find $a_0 = 68$ and 88\,THz\,amu$^{1/2}$ and $a_1 = 4.7$ and 3.9\,amu$^{1/2}$ for the $\nu_2=0$ and $\nu_2=1$ inversion modes, respectively. For \ce{^{14}NH3}, this results in sensitivity coefficients $K_\mu^\text{inv}=-4.2$ and $-3.6$.

Alternatively, an expression for the sensitivity coefficients may be obtained from the derivative of Eq.~\eqref{eq:WKBsplitting}. By explicitly taking the $\mu$ dependence of the vibrational energy term in the exponent of Eq.~\eqref{eq:WKBsplitting} into account, Flambaum and Kozlov derived\cite{FlambaumKozlov2007NH3}

\begin{equation}
K_\mu^\text{inv} =-\frac{1}{2}\left (1 + G + \frac{G}{2}\frac{\omega_0}{\Delta U-\tfrac{1}{2}\omega_0} \right ).
\end{equation}

\noindent
This expression yields $K_\mu^\text{inv}=-4.4$ and $-3.4$ respectively, in fair agreement with the result obtained from the fit through the isotopologue data.

\begin{figure}[bt]
\centering
\includegraphics[width=0.9\columnwidth]{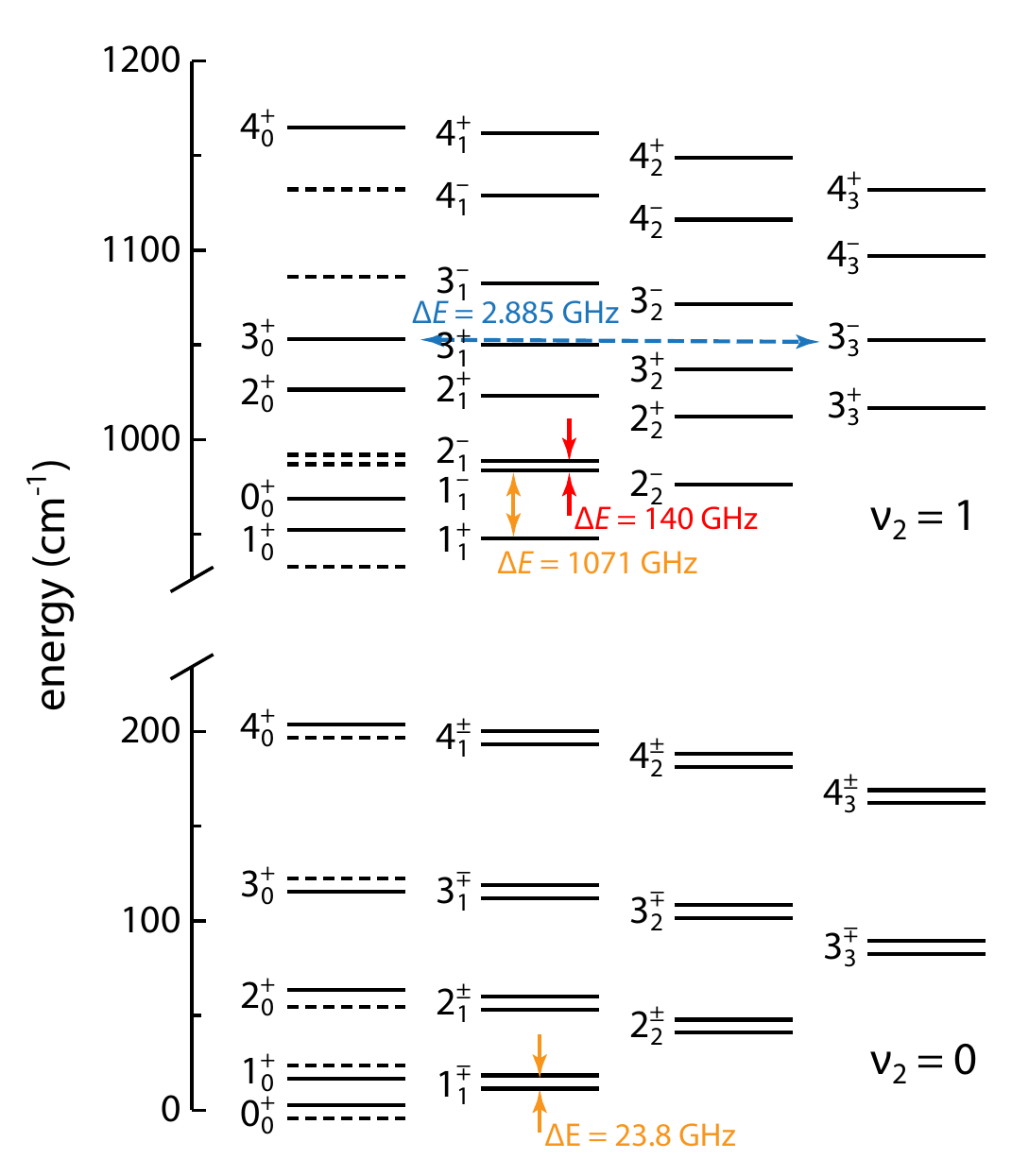}
\caption{Level diagram of the lower rotational energy levels of the $\nu_2=0$ and $\nu_2=1$ states of \ce{NH3}. Each level is characterized by the rotational quantum numbers $J_K$ and the symmetry label, $+/-$, of the rovibronic wave function. The inversion doubling in the $\nu_2=0$ state has been exaggerated for clarity. The dashed lines indicate symmetry-forbidden levels.
\label{fig:ammonia_Kladders}}
\end{figure}

Astronomical observations of the inversion splitting of \ce{NH3}, redshifted to the radio range of the electromagnetic spectrum, led to stringent constraints at the level of $(-3.5\pm 1.2)\times 10^{-7}$ at $z=0.69$\cite{Kanekar2011} and $(0.8\pm 4.7)\times 10^{-7}$ at $z=0.89$\cite{Henkel2009}. These constraints were derived by comparing the inversion lines of ammonia with pure rotation lines of \ce{HC3N}\cite{Henkel2009} and \ce{CS} and \ce{H2CO}\cite{Kanekar2011} and rely on the assumption that these different molecular species reside at the same redshift.

The relatively high sensitivity of the inversion frequency in ammonia also allows for a test of the time independence of $\mu$ in the current epoch. A molecular fountain based on a Stark-decelerated beam of ammonia molecules has been suggested as a novel instrument to perform such measurement~\cite{Bethlem2008}. By comparing the inversion splitting with an appropriate frequency standard $\Delta\mu/\mu$ can be constrained or a possible drift may be detected.

\subsection{Near degeneracies between inversion and rotation energy}
\label{sec:inversion_degeneracies}

The sensitivity coefficients for the inversion frequency in the different isotopologues of ammonia are two orders of magnitude larger than those found for rovibronic transitions in molecular hydrogen and carbon monoxide and one order of magnitude larger than a pure vibrational transition. Yet, Eq.~\eqref{eq:toy} predicts even higher sensitivities if the inversion splitting becomes comparable to the rotational splitting, as this may introduce accidental degeneracies. Such degeneracies do not occur in the vibrational ground state of ammonia, but may happen in excited $\nu_2$ vibrational states. In Fig.~\ref{fig:ammonia_Kladders}, a rotational energy diagram of ammonia in the $\nu_2=0$ and $\nu_2=1$ state is shown. As can be seen in this figure, the larger inversion splitting in the $\nu_2=1$ state results in smaller energy differences between different rotational states within each $K$ manifold. This is in particular the case for the $J_K^s=1_1^-$ and $2_1^-$ levels that have an energy difference of only 140\,GHz. Using Eq.~\eqref{eq:toy} we find $K_\mu = 18.8$ for this inversion-rotation transition. As \ce{NH3} and \ce{ND3} are symmetric top molecules, transitions that have $\Delta K\neq 0$ are not allowed and this reduces the number of possible accidental degeneracies. Kozlov \emph{et. al}\cite{Kozlov2010} investigated transitions in the $\nu_2=0$ state of asymmetric isotopologues of ammonia (\ce{NH2D}, \ce{ND2H}), in which transitions with $\Delta K \neq 0$ are allowed, but found no sensitive transitions, mainly because the inversion splitting in the $\nu_2=0$ mode is much smaller than the rotational splitting.

It is interesting to note that ``forbidden'' transitions with $\Delta K=\pm 3$ gain amplitude in the $\nu_2=1$ state of \ce{NH3} due to perturbative mixing of the (accidental) near-degenerate $J_K^s=3_0^+$ and $3_3^-$ levels~\cite{Laughton1976}. Using Eq.~\eqref{eq:toy} to estimate the sensitivity coefficient of the 2.9\,GHz transition between these two levels, we find $K_\mu=-938$. However, since these levels both have positive overall parity, a two-photon transition is required to measure this transition directly.

The hydronium ion (\ce{H3O+}) has a similar structure to ammonia but experiences a much smaller barrier to inversion. As a consequence the inversion splitting in the ground vibrational state in hydronium is much larger than for ammonia. Kozlov and Levshakov\cite{KozlovLevshakov2011} found that pure inversion transitions in hydronium have a sensitivity of $K_\mu^\text{inv}=-2.5$ and, in addition, identified several mixed transitions with sensitivity coefficients ranging from $K_\mu=-9.0$ to $+5.7$. Mixed transitions in the asymmetric hydronium isotopologues H$_2$DO$^+$ and D$_2$HO$^+$ possess sensitivity coefficients ranging from $K_\mu=-219$ to $+11$\cite{Kozlov2011}.

\subsection{Internal rotation; from methanol to methylamine\label{sec:internal_rotation}}
While inversion doublets of ammonia-like molecules exhibit large sensitivity coefficients, even larger sensitivity coefficients arise for molecules that exhibit internally hindered rotation, in which one part of a molecule rotates with respect to the remainder. This is another example of a classically-forbidden tunneling motion that is frequently encountered in polyatomic molecules.
This subject of the interaction between such hindered rotation, also referred to as torsion,
and its quantum mechanical description has been investigated since the 1950s.\cite{Kivelson1954,LinSwalen1959,Herschbach1959,Kirtman1962,Lees1968,Lees1973}

In this section we outline the procedure for obtaining the sensitivity coefficients in internal rotor molecules containing a $C_{3v}$ symmetry group and show that a particular combination of molecular parameters can be identified that results in the highest sensitivity coefficients. The fact that methanol possesses transitions with enhanced sensitivity coefficients was discovered independently by Jansen \emph{et al.}\cite{Jansen2011PRL} and by Levshakov \emph{et al.}\cite{LevshakovKozlov2011}\\

\begin{figure}[bt]
\centering
\includegraphics[width=1\columnwidth]{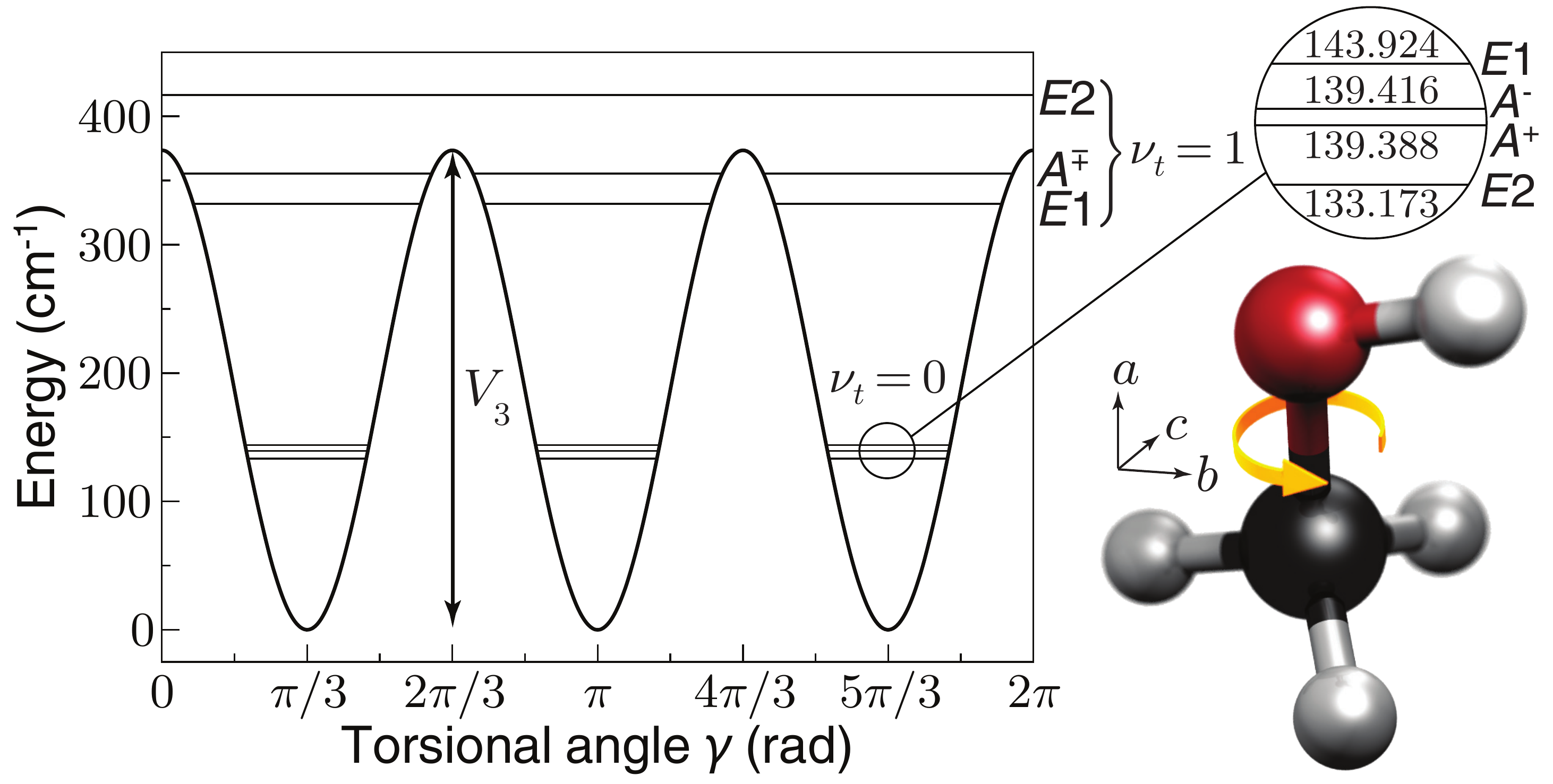}
\caption{Variation of the potential energy of methanol with the relative rotation of the \ce{OH} group with respect to the \ce{CH3} group and a schematic representation of the molecule. Shown are the $J=1$, $|K|=1$ energies of the lowest torsion-vibrational levels.
\label{fig:methanol_potential}}
\end{figure}

One of the simplest molecules that exhibits hindered internal rotation is methanol (\ce{CH3OH}). Methanol, schematically depicted on the right-hand side of Fig.~\ref{fig:methanol_potential}, consists of a methyl group (\ce{CH3}) with a hydroxyl group (\ce{OH}) attached. The overall rotation of the molecule is described by three rotational constants $A$, $B$, and $C$, associated with the moments of inertia $I_a$, $I_b$, and $I_c$, respectively, along the three principal axes of the molecule. The total angular momentum of the molecule is given by the quantum number $J$, while the projection of $J$ onto the molecule fixed axis is given by $K$.

In addition to the overall rotation, the flexible \ce{CO} bond allows the methyl group to rotate with respect to the hydroxyl group, denoted by the relative angle $\gamma$. This internal rotation is not free but hindered by a threefold potential barrier,\cite{Swalen1955} shown on the left-hand side of Fig.~\ref{fig:methanol_potential}, with minima and maxima that correspond to the staggered and eclipsed configuration of the molecule, respectively. The vibrational levels in this well are denoted by $\nu_t$.

When we neglect the slight asymmetry of the molecule as well as higher-order terms in the potential and centrifugal distortions, the lowest-order Hamiltonian can be written as

\begin{multline}
H = \frac{1}{2} \frac{P_a^2}{I_a} + \frac{1}{2} \frac{P_b^2}{I_b} + \frac{1}{2} \frac{P_c^2}{I_c} + \frac{1}{2} \frac{1}{I_\text{red}} p_{\gamma}^2 + \frac{1}{2} V_3 (1 - \cos {3\gamma}),\\
\text{ with} \quad I_\text{red} = \frac{I_{a1} I_{a2}}{I_a}.
\label{eq:H_IR}
\end{multline}

\noindent
The first three terms describe the overall rotation around the $a$, $b$ and $c$ axis, respectively. The fourth term describes the internal rotation around the $a$ axis, with $I_{red}$ the reduced moment of inertia along the $a$-axis, $I_{a2}$ the moment of inertia of the methyl group along its own symmetry axis and $I_{a1}$ the part of $I_a$ that is attributed to the \ce{OH} group; $I_{a1} = I_a - I_{a2}$. Note that in the derivation of Eq.~(\ref{eq:H_IR}) an axis transformation was applied in order to remove the coupling between internal and overall rotation. The fifth term is the lowest order term arising from the torsional potential. If the potential were infinitely high, the threefold barrier would result in three separate harmonic potentials, whereas the absence of the potential barrier would result in doubly degenerate free-rotor energy levels. In the case of a finite barrier, quantum-mechanical tunneling mixes the levels in different wells of the potential. As a result, each rotational level is split into three levels of different torsional symmetry, labeled as $A$, $E1$, or $E2$. Following Lees~\cite{Lees1973}, $E1$ and $E2$-symmetries are labeled by the sign of $K$; i.e, levels with $E1$-symmetry are denoted by a positive $K$-value, whereas levels with $E2$-symmetry are denoted by a negative $K$-value. For $K\neq 0$, $A$ levels are further split into $+/-$ components by molecular asymmetry. For $K = 0$, only single $E$ and $A^{+}$ levels exist.

The splitting between the different symmetry levels is related to the tunneling frequency between the different torsional potential wells and is therefore very sensitive to the reduced moment of inertia, similar to the inversion of the ammonia molecule. It was shown by Jansen \emph{et al.}\cite{Jansen2011PRL,Jansen2011} that a pure torsional transition in methanol has a sensitivity coefficient of $K_\mu=-2.5$. However, pure torsional transitions are forbidden, since they possess a different torsional symmetry. Sensitivity coefficients for allowed transitions in methanol and other internal rotor molecules can be obtained by calculating the level energies as a function of $\mu$ and taking the numerical derivative, in accordance with Eq.~(\ref{eq:Kmu}). This can be achieved by scaling the different parameters in the molecular Hamiltonian according to their $\mu$ dependence. The physical interpretation of the lowest-order constants is straightforward and the scaling relations can be derived unambiguously. Higher order parameters pose a problem since their physical interpretation is not always clear. Jansen \emph{et al.}\cite{Jansen2011PRL,Jansen2011} derived the scaling relations for these higher-order constants by considering them as effective products of lower-order torsional and rotational operators. Ilyushin \emph{et al.} showed that the scaling of the higher order constants only contributes marginally to the sensitivity coefficient of a transition.~\cite{Ilyushin2012}

Jansen \emph{et al.}\cite{Jansen2011PRL,Jansen2011} employed the state-of-the art effective Hamiltonian that is implemented in the {\sc belgi} code\cite{Hougen1994} together with a set of 119 molecular parameters.\cite{Xu1999,Xu2008}
Similar calculations were performed by Levshakov \emph{et al.} using a simpler model containing only six molecular parameters.\cite{LevshakovKozlov2011}
The two results are in excellent agreement and sensitivity coefficients for transitions in methanol range from $-42$ for the $5_1\rightarrow 6_0 A^+$ transition at 6.6\,GHz to $+53$ for the $5_2\rightarrow 4_3 A^+$ transition at 10.0\,GHz.

\begin{figure}[bt]
\centering
\includegraphics[width=1\columnwidth]{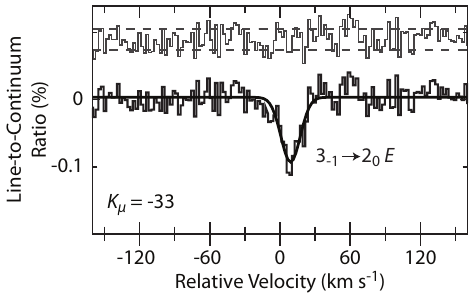}
\caption{Observed spectrum of the $3_{-1} - 2_0E$ methanol transition observed in the gravitational lensed object PKS1830-211 with the Effelsberg radio telescope.\cite{Bagdonaite2013}
\label{fig:methanolobs}}
\end{figure}

The large number of both positive and negative sensitivity coefficients makes methanol a preferred target system for probing a possible variation of $\mu$, since this makes it possible to test variation of $\mu$ using transitions in a single molecular species, thereby avoiding the many systematic effects that plague tests that are based on comparing transitions in different molecules. Following the recent detection of methanol in the gravitationally lensed object PKS1830-211 (PKS referring to the Parkes catalog of celestial objects, with 1830 and -211 referring to RA and dec coordinates as for quasars; the PKS1830-211 system is a radio-loud quasar at $z_{\rm{emis}}=2.51$) in an absorbing galaxy at a redshift of $z_{\rm{abs}}=0.89$~\cite{Muller2011}, Bagdonaite~\emph{et al.}\cite{Bagdonaite2013} used four transitions that were observed in this system using the 100m radio telescope in Effelsberg  to constrain $\Delta\mu/\mu$ at $(0.0\pm 1.0)\times 10^{-7}$ at a look-back time of 7 billion years. A spectrum of the $3_{-1} - 2_0E$ methanol line, the line with the largest sensitivity to $\mu$-variation observed at high redshift, is shown in Fig.~\ref{fig:methanolobs}.

The enhancements discussed in methanol, generally occur in any molecule that contains an internal rotor with $C_{3v}$ symmetry.
Jansen~\emph{et. al} constructed a simple model that predicts whether a molecule with such $C_{3v}$ group is likely to have large sensitivity coefficients.\cite{Jansen2011PRL} This "toy" model decomposes the energy of the molecule into a pure rotational and a pure torsional part, \emph{cf.} Eq.~\eqref{eq:toy}. The rotational part is approximated by the well-known expression for the rotational energy levels of a slightly asymmetric top

\begin{equation}
E_\text{rot}(J,K)=\frac{1}{2}\left (B+C \right )J\left (J+1 \right )+\left (A-\frac{B+C}{2} \right )K^2,
\end{equation}

\noindent
with $A$, $B$, and $C$ the rotational constants along the $a$, $b$, and $c$ axis of the molecule, respectively. The torsional energy contribution is approximated by a Fourier expansion as~\cite{LinSwalen1959}

\begin{equation}
E_\text{tors}(K)=F \left [a_0+a_1\cos\left \{ \frac{2\pi}{3}\left (\rho K +\sigma \right )\right \} \right ],
\label{eq:Etors}
\end{equation}

\begin{figure}[bt]
\centering
\includegraphics[width=1\columnwidth]{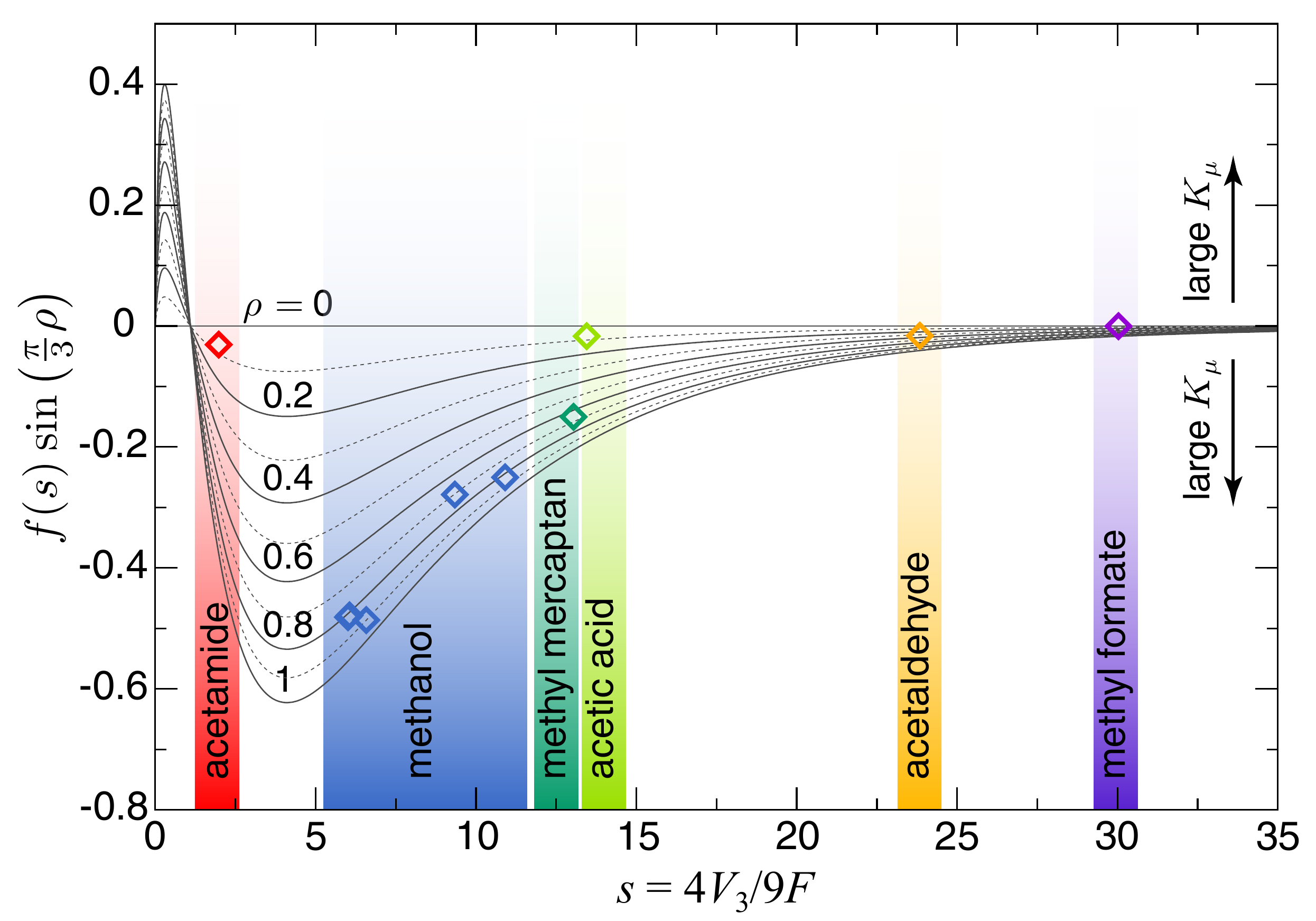}
\caption{The product $f(s)\sin\left (\tfrac{\pi}{3}\rho \right )$ which is a measure of the maximum value of $K_\mu$ (see text). Also shown are data points for molecules containing a internal rotor with $C_{3v}$ symmetry for which the sensitivity coefficients have been calculated.
\label{fig:toy_model}}
\end{figure}

\noindent
where $F\simeq\frac{1}{2}\hbar^2 I_{\text{red}}^{-1}$ is the constant of the internal rotation, $\rho\simeq I_{a2}/I_a$ is a dimensionless constant reflecting the coupling between internal and overall rotation, and $\sigma=0,\pm 1$ is a constant relating to the torsional symmetry. The expansion coefficients $a_0$ and $a_1$ depend on the shape of the torsional potential. Since we are mainly interested in the torsional energy difference, $a_0$ cancels, and $a_1$ is obtained from

\begin{equation}
a_1 = A_1s^{B_1}e^{-C_1\sqrt{s}},
\end{equation}

\noindent
with $A_1=-5.296$, $B_1=1.111$, and $C_1=2.120$~\cite{Jansen2011}. The dimensionless parameter $s=4V_3/9F$, with $V_3$ the height of the barrier, is a measure of the effective potential. The sensitivity of a pure torsional transition is given by $K_\mu^\text{tors}=(B_1-1)-\tfrac{1}{2}C_1\sqrt{s}$. Inserting the different terms in Eq.~\eqref{eq:toy} reveals that the sensitivity coefficient of a transition is roughly proportional to $f(s)\sin\left (\tfrac{\pi}{3}\rho \right )$, with $f(s)=-2a_1\left (K_\mu^\text{tors}+1 \right )$. This function is plotted in Fig.~\ref{fig:toy_model} for several values of $\rho$. The curves can be regarded as the maximum sensitivity one may hope to find in a molecule with a certain $F$ and transition energy $h\nu$. The maximum sensitivity peaks at $s=4$ and $\rho=1$. From the figure it is seen that only methanol, and to a lesser extend methyl mercaptan, lie close to this maximum. Indeed, the highest sensitivities are found in these molecules. It is unlikely that other molecules are more sensitive than methanol since the requirement for a large value of $\rho$ and a relatively low effective barrier favors light molecules. \\

\begingroup
\squeezetable
\centering
\renewcommand{\arraystretch}{1.25}
\begin{table*}[bt!]
\caption{Selection of diatomic and polyatomic molecules that are used or have been proposed to probe a possible variation of $\mu$. The second and third columns list the relevant electronic state and the different types of energies involved in the transitions, while the fourth and fifth column give the range of sensitivities and corresponding bibliographical references.  \label{tab:moleculelist}}
\begin{ruledtabular}
\begin{tabularx}{\textwidth}{l l l c c}
                    & electronic state      & origin                                                  & $K_\mu$                & Ref. \\ \hline
Diatomic molecules  &      &                                                         &                        &       \\
~~~~~~\ce{H2} & $B ^1\Sigma^+_u \leftarrow X ^1\Sigma^+_g/C ^1\Pi_u \leftarrow X ^1\Sigma^+_g$ & $E_\text{el}/ E_\text{vib}$  & $-0.054<K_\mu<+0.019$    & [\!\!\citenum{Meshkov2006,Ubachs2007}]\\
~~~~~~\ce{HD} & $B ^1\Sigma^+_u \leftarrow X ^1\Sigma^+_g/C ^1\Pi_u \leftarrow X ^1\Sigma^+_g$            & $E_\text{el}/ E_\text{vib}$  & $-0.052<K_\mu<+0.012$    & [\onlinecite{Ivanov2008}]\\
~~~~~~\ce{CH} & $X ^2\Pi$           & $E_\text{fs}/E_\text{rot}/E_\Lambda$                             &  $-6.2<K_\mu<+2.7$ & [\!\!\citenum{Kozlov2009,DeNijs2012}]     \\
~~~~~~\ce{CD} & $X ^2\Pi$          & $E_\text{fs}/E_\text{rot}/E_\Lambda$ &   $-67<K_\mu<+18$   &   [\!\!\citenum{DeNijs2012}]\\
~~~~~~\ce{OH} & $X ^2\Pi$           & $E_\text{fs}/E_\text{rot}/E_\Lambda$ &   $-460<K_\mu<-0.50$   &   [\!\!\citenum{Kozlov2009}]\\
~~~~~~\ce{NO} & $X ^2\Pi$           & $E_\text{fs}/E_\text{rot}/E_\Lambda$ &   $-38.9<K_\mu<+6.81$   &   [\!\!\citenum{Kozlov2009}]\\
~~~~~~\ce{LiO} & $X ^2\Pi$           & $E_\text{fs}/E_\text{rot}/E_\Lambda$ &   $-4.24<K_\mu<-0.95$   &   [\!\!\citenum{Kozlov2009}]\\
~~~~~~\ce{NH+} & $a ^4\Sigma^- \leftarrow X ^2\Pi \,(\nu=0,1)$          & $E_\text{el}/E_\text{rot}$                             & $-185.8 < K_\mu < +126.9$                         &   [\onlinecite{Beloy2011}]                       \\
~~~~~~\ce{CO} &$A ^1\Pi \leftarrow X ^1\Sigma^+$        & $E_\text{el}/ E_\text{vib}$  & $-0.071<K_\mu<+0.003$    & [\onlinecite{Salumbides2012}]                   \\
              & $a ^3\Pi$ & $E_\text{fs}/ E_\text{rot}$ & $-334<K_\mu<+128$        & [\!\!\citenum{BethlemUbachs2009,DeNijs2011}]\\
Polyatomic molecules   &   &                                            &                        &\\
~~~~~~\ce{NH3}         & $\tilde{X}$  & $E_\text{inv}$                             & $-4.2$                 & [\!\!\citenum{FlambaumKozlov2007NH3}]\\
~~~~~~\ce{ND3}         & $\tilde{X}$  & $E_\text{inv}$                             & $-5.6$                 &
[\!\!\citenum{vanVeldhoven2004}]\\
~~~~~~\ce{NH2D}/\ce{ND2H}  & $\tilde{X}$   & $E_\text{inv}/ E_\text{rot}$               & $-1.54<K_\mu<+0.10$                       & [\!\!\citenum{Kozlov2010}]\\
~~~~~~\ce{H3O+}        & $\tilde{X}$  & $E_\text{inv}$                             & $-2.5$                 & [\!\!\citenum{KozlovLevshakov2011}]\\
                       &  $\tilde{X}$  & $E_\text{inv}/ E_\text{rot}$               & $-9.0<K_\mu<+5.7$    & [\!\!\citenum{KozlovLevshakov2011}]\\
~~~~~~\ce{H2DO+}/\ce{D2HO+} &  $\tilde{X}$  & $E_\text{inv}/ E_\text{rot}$           & $-219<K_\mu<+11.0$  & [\!\!\citenum{Kozlov2011}]\\
~~~~~~\ce{H2O2}         & $\tilde{X}$ & $E_\text{inv}/E_\text{rot}$ &  $-36.5<K_\mu<+13.0$  &[\onlinecite{Kozlov2011b}]\\
~~~~~~\ce{CH3OH}        & $\tilde{X}$  & $E_\text{tors}/ E_\text{rot}$              & $-88<K_\mu<+330$  &[\!\!\citenum{Jansen2011PRL,LevshakovKozlov2011,Jansen2011}]\\
~~~~~~\ce{CH3SH}        & $\tilde{X}$  & $E_\text{tors}/ E_\text{rot}$               & $-14.8<K_\mu<+12.2$  &[\onlinecite{Jansen2013}]\\
~~~~~~\ce{CH3COH}       & $\tilde{X}$  & $E_\text{tors}/ E_\text{rot}$              & $-3.7 < K_\mu < -0.5$ & [\onlinecite{Jansen2011}]\\
~~~~~~\ce{CH3CONH2}     & $\tilde{X}$  & $E_\text{tors}/ E_\text{rot}$              & $-1.34 < K_\mu < +0.06$ & [\onlinecite{Jansen2011}]\\
~~~~~~\ce{HCOOCH3}      & $\tilde{X}$  & $E_\text{tors}/ E_\text{rot}$ & $-1.07<K_\mu<-0.03$ & [\onlinecite{Jansen2011}]\\
~~~~~~\ce{CH3COOH}      &$\tilde{X}$  & $E_\text{tors}/ E_\text{rot}$ & $-1.36<K_\mu<-0.27$ & [\onlinecite{Jansen2011}]\\
~~~~~~\textit{l-}\ce{C3H}& $\tilde{X} ^2\Pi$ & $E_\text{RT}/ E_\text{vib}/ E_\text{rot}$  & $-19<K_\mu<+742$  &[\onlinecite{Kozlov2013}]\\
~~~~~~\ce{CH3NH2}        & $\tilde{X}$ & $E_\text{inv}/E_\text{tors}/E_\text{rot}$  & $-19<K_\mu<+24$ &[\onlinecite{Ilyushin2012}]\\

\end{tabularx}
\end{ruledtabular}
\end{table*}
\endgroup

\begin{figure}[bt]
\centering
\includegraphics[width=1\columnwidth]{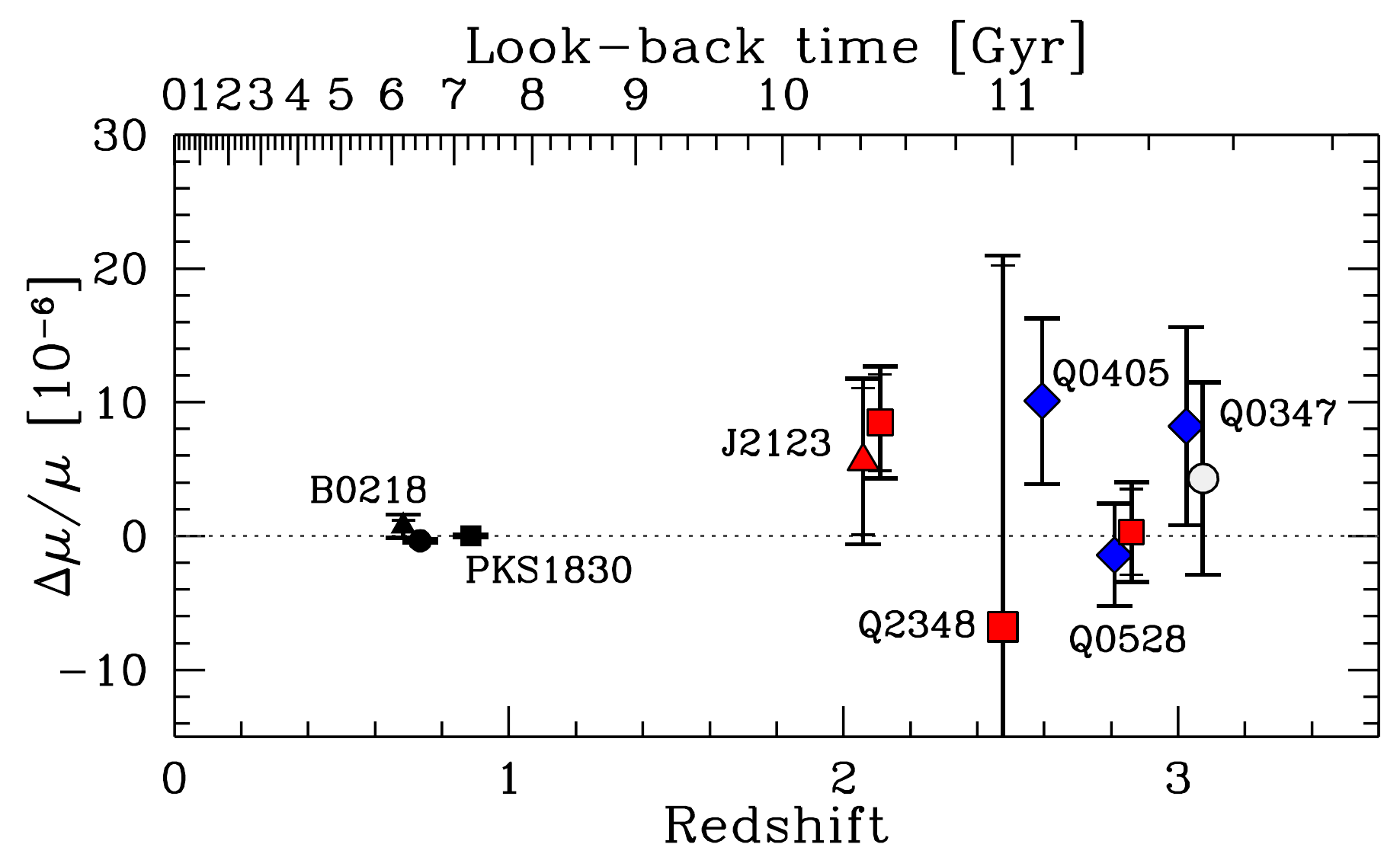}
\caption{Current astrophysical constraints on $\Delta\mu/\mu$ based on \ce{H2}, \ce{NH3}, and \ce{CH3OH} data. The constraints at higher redshift were derived from optical transitions of \ce{H2} in the line of sight of 5 different quasars (Q0528-250\cite{King2008,King2011}, Q2123-0050\cite{Malec2010,vanWeerdenburg2011}, Q0347-383\cite{King2008,WendtMolaro2012}, Q2348-011\cite{Bagdonaite2012}, and Q0405-443\cite{King2008}) and typically yield $\Delta\mu/\mu \lesssim 10^{-5}$. At intermediate redshift, the most stringent tests are based on microwave and radio-frequency transitions in methanol (PKS1830-211\cite{Bagdonaite2013}) and ammonia (B0218+357\cite{Murphy2008,Kanekar2011} and PKS1830-211\cite{Henkel2009}) and constrain $\Delta\mu/\mu$ at the $10^{-7}$ level.
\label{fig:mu_vs_z}}
\end{figure}

We have seen that molecules that undergo inversion or internal rotation may possess transitions that are extremely sensitive to a possible variation of $\mu$. A molecule that exhibits both types of these motions, and has also been observed in PKS1830-211\cite{Muller2011}, is methylamine (\ce{CH3NH2}); hindered internal rotation of the methyl (\ce{CH3}) group with respect to the amino group (\ce{NH2}), and tunneling associated with wagging of the amino group\cite{Tsuboi1964}. The coupling between the internal rotation and overall rotation in methylamine is rather strong resulting in a large value of $\rho$, which is favorable for obtaining large enhancements of the sensitivity coefficients. Ilyushin \emph{et. al}\cite{Ilyushin2012} have calculated sensitivity coefficients for many transitions in methylamine and found that the transitions can be grouped in pure rotation transitions with $K_\mu=-1$, pure inversion transitions with $K_\mu\approx-5$, and mixed transitions with $K_\mu$ ranging from $-19$ to $+24$.

\section{Summary and outlook\label{sec:summary}}
In this paper we discussed several molecular species that are currently being used in studies aimed at constraining or detecting a possible variation of the proton-to-electron mass ratio. These molecules, together with a range of other species of relevance for $\mu$-variation, are listed in Table~\ref{tab:moleculelist}. From this table it can be seen that the highest sensitivities are found in open-shell free radicals and polyatomics, due to the systematic occurrence of near-degenerate energy levels in these molecules. Several of these molecules have been observed already at high redshift, others have been observed in the interstellar medium of our local galaxy providing a prospect to be observed at high redshift in the future, whereas other molecules, in particular low-abundant isotopic species might be suitable systems for tests of $\mu$ variation in the present epoch.

Astrophysical and laboratory studies are complementary as they probe $\mu$ variation at different time scales. The most stringent constraint in the current epoch sets $\Delta\mu/\mu<6\times 10^{-14}$\,yr$^{-1}$ and was obtained from comparing rovibrational transitions in \ce{SF6} with a \ce{Cs} fountain clock.\cite{Shelkovnikov2008}
On a cosmological time scale, at the highest redshifts observable molecular hydrogen remains the target species of choice limiting a cosmological variation of $\mu$ below $|\Delta\mu/\mu|<1\times 10^{-5}$.\cite{Malec2010,vanWeerdenburg2011} Current constraints derived from astrophysical data are summarized graphically in Fig.~\ref{fig:mu_vs_z}.
At somewhat lower redshifts ($z\sim 1$) constraints were derived from highly sensitive transitions in ammonia and methanol probed by radio astronomy are now producing limits on a varying $\mu$ of  $|\Delta\mu/\mu| < 10^{-7}$.\cite{Muller2011,Henkel2009,Kanekar2011,Bagdonaite2013}

This result, obtained from observation of methanol at redshift $z=0.89$, represents the most stringent bound on a varying constant found so far.\cite{Bagdonaite2013,Bagdonaite2013a}
Its redshift corresponds to a look-back time of 7 Gyrs (half the age of the Universe), and it translates into $\dot{\mu}/\mu = (1.4 \pm 1.4) \times 10^{-17}$/yr if a linear rate of change is assumed.
As it is likely that $\mu$ changes faster or at the same rate as $\alpha$, \emph{cf.} Eq.~(\ref{GUT}), this result is even more constraining than the bounds on varying constants obtained with optical clocks in the laboratory.\cite{Rosenband2008}

\begin{acknowledgments}
We thank Julija Bagdonaite, Adrian de Nijs, and Edcel Salumbides (VU Amsterdam) as well as Julian Berengut (UNSW Sydney) for helpful discussions. This research has been supported by the FOM-program `Broken Mirrors \& Drifting Constants'. P.~J. and W.~U. acknowledge financial support from the Templeton Foundation. H.~L.~B acknowledges financial support from NWO via a VIDI-grant and from the ERC via a Starting Grant.
\end{acknowledgments}


\begin{thebibliography}{107}
\bibitem [{Foo()}]{FootnoteA}%
  \BibitemOpen
  \href@noop {} {\ }\bibinfo {note} {Note that spectroscopic transitions in
  molecules probe the inertial or kinematic mass, as it enters in the
  Schr$\rm{\ddot{o}}$dinger equation, rather than a gravitational
  mass.}\BibitemShut {Stop}%
\end{thebibliography}
%

\end{document}